\documentclass[a4paper,12pt]{article}
\usepackage[pctex32]{graphics}
\usepackage{amssymb,amsmath}
\usepackage{amsmath,amssymb}
\usepackage{latexsym}
\usepackage{epsfig}
\usepackage[english]{babel}

\newcommand{\be}{\begin{equation}}
\newcommand{\ee}{\end{equation}}
\newcommand{\ba}{\begin{eqnarray}}
\newcommand{\ea}{\end{eqnarray}}

\begin{document}

\begin{titlepage}

\vspace{5mm}

\begin{center}

{\Large \bf Renormalizability,  vDVZ discontinuity  and Newtonian singularity   in higher-derivative gravity}

\vskip .6cm

\centerline{\large
 Yun Soo Myung$^{a}$}

\vskip .6cm

{Institute of Basic Science and Department of Computer Simulation,
\\Inje University, Gimhae 50834, Korea \\}

\end{center}

\begin{center}
\underline{Abstract}
\end{center}
It was proposed that  if a higher-derivative gravity is renormalizable, it implies  necessarily  a finite Newtonian potential at the origin,
but the reverse of this statement is not true. Here we show that the reverse is true when taking into account the vDVZ discontinuity which states
that the theory obtained from massive one by taking zero mass limit is not equivalent to the theory obtained in the zero mass case.
The surviving degree of freedom in the zero mass limit is an extra scalar  which does not affect the light bending angle, but affects the Newtonian potential.
This asserts that in order to make the singularity cancellation, the number of massive ghost and healthy  tensors  matches with that of  massive ghost and healthy scalars.

 \vskip .6cm

\vskip 0.8cm

\vspace{15pt} \baselineskip=18pt

\noindent $^a$ysmyung@inje.ac.kr \\

\thispagestyle{empty}
\end{titlepage}

\newpage
\section{Introduction}
It was shown   that the renormalizability in higher-derivative gravity might be related to the behavior of the classical potential of the model.
Explicitly, there is a conjecture that renormalizable higher-derivative gravity has  a finite Newtonian potential at the  origin~\cite{Accioly:2013hwa,Modesto:2014eta}.
This relation was first notified in Stelle's seminal work~\cite{Stelle:1976gc} which showed that the fourth-derivative gravity is renormalizable, nonunitary and has a finite potential at origin.
In this case, a massive ghost tensor  and a massive healthy scalar  contribute in such a manner to cancel out the Newtonian singularity of a massless tensor.
Recently, it was  conjectured that the reverse of the above statement is not true, which indicates that the finiteness of the Newtonian potential at the origin
is a necessary but not sufficient condition for the renormalizability of the model~\cite{Asorey:1996hz,Giacchini:2016xns}. The model used in~\cite{Giacchini:2016xns,Accioly:2017rft}
includes  a massive ghost   tensor,  massive ghost and heathy scalars  in addition to  a healthy tensor. Even though the potential is finite at $r=0$, it is non-renormalizable by power counting. Actually, two massive (ghost and healthy) scalars make a contribution 1/3 to the massless and massive ghost tensors ($1-4/3$).
This is not the case that   the number of massive ghost (healthy) tensors matches with the number of massive healthy (ghost)  scalars.

On the other hand, it was known that Fierz and Pauli (FP) in 1939 have  obtained 5 propagating degrees of freedom (DOF) of a massive tensor
by adding a  mass term of $m^2(h_{\mu\nu}h^{\mu\nu}-h^2)/2$ to the bilinearized Einstein-Hibert action~\cite{Fierz:1939ix}. It is well known  that a massless tensor  has 2 DOF.
An inevitable mismatch between  massive and massless cases was first realized in 1970 by van Dam and Veltman~\cite{vanDam:1970vg} and independently by
Zakharov~\cite{Zakharov:1970cc}. Then, it is known as the vDVZ discontinuity which states that  the theory obtained from  massive one by taking zero mass limit ($m\to0$) is not equivalent to the theory obtained in the zero mass case ($m=0$).
Especially, the former has   3 DOF while the latter has 2 DOF  which shows a difference of 1 DOF. When using the Stueckelberg formalism~\cite{Stueckelberg:1957zz}, one may find  the origin of  the vDVZ discontinuity~\cite{Hinterbichler:2011tt}.
After applying  this formalism to the FP massive gravity action, a scalar field which was introduced by  Stueckelberg  to maintain the gauge symmetry, was coupled to the external  source.
 The coupling between the source and the  Stueckelberg field was identified
as the origin of the  vDVZ discontinuity.
This Stueckelberg scalar behaves as an attractive force in the theory and affects the Newtonian potential, but not the light bending angle. We note that the Newtonian potential of $V_{m=0}(r)=-GM/r$
differs from $V_{m\to0}(r)=-\frac{4}{3}GM/r$ in the zero mass limit of $e^{-mr}\approx1$.
Clearly, the scalar causes the mismatch ($-1/3$) in the Newtonian potential between massless and massive gravities.

Hence, we propose that  the vDVZ discontinuity is  related closely  to the  the singularity cancellation of the Newtonian potential at the origin.

In this work, we explore why the matching of the number of ghost and  healthy  modes between the spin-2 and spin-0 massive sectors is necessary to make the singularity cancellation at the origin.
This will be explained by introducing the vDVZ discontinuity appeared  in the zero mass limit of the massive gravity.

The organization of our work is as follows. In section 2, we study the fourth-derivative gravity as a toy model of higher derivative gravities.
We will explain the singularity cancellation by introducing the zero mass limit where  the vDVZ discontinuity occurs.
This model shows that the theory without any kind of non-locality could be free from the Newtonian singularity.
Section 3 is devoted to explaining the finiteness of Newtonian potential obtained from a full sixth-derivative gravity by taking the small mass limit.
We present in section 4 that the half sixth-derivative gravity is not a suitable model which explains the connection between the finiteness of the potential and renormalizability.
This is because this model lacks for the matching of the number of healthy and ghost modes between spin-2 and spin-0 massive sectors  needed to implement a singularity cancellation.
In section 5, we introduce a polynomial form of infinite-derivative gravity to explain the singularity cancellation by making use of the vDVZ discontinuity.
Finally, we discuss our results in section 6.

\section{Fourth-derivative theory of gravity}
 The fourth-derivative gravity in four dimensions is defined by
 \begin{equation} \label{action1}
S_{\rm 4th}=\int d^4x \sqrt{-g}\Big[\frac{2}{\kappa^2} R+\frac{\alpha}{2}R^2+\frac{\beta}{2}R_{\mu\nu}^2\Big]
 \end{equation}
 with $\kappa^2=4\kappa_4(\kappa_4=8\pi G)$. This action was first  employed to prove the renormalizability of fourth-derivative  gravity~\cite{Stelle:1976gc} and hence, it is considered as a prototype of higher-derivative gravities. A key point is that the last two terms  are necessary to  achieve the renormalizability,  but the last term induces the ghost state.
 We have seen that the fourth-derivative gravity is not a healthy theory because of a massive ghost tensor that violates the unitarity at the tree level.
 Recently, it was confirmed that renormalizable higher-derivative gravities are nonunitary~\cite{Accioly:2017rft}. Hence, the ghost problem  is not a relevant issue  in this work.
 We are interested in exploring the connection between  renormalizability of the theory and finiteness  of the Newtonian potential at the origin.

    In order to find the Newtonian potential, one has first  to compute the propagator. For this purpose,
     we expand the metric tensor  $g_{\mu\nu}= \eta_{\mu\nu}+\kappa h_{\mu\nu}$  around the Minkowski metric $\eta_{\mu\nu}={\rm diag}(+,-,-,-)$.
Bilinearizing the Lagrangian in Eq. (\ref{action1})  together with  imposing the de Donder gauge of ${\cal L}_{\rm gf}=-(\partial^\mu h_{\mu\nu}-\partial_\nu h/2)^2/2\lambda$, one obtains ${\cal L}^{\rm bil}_{\rm 4th}=h_{\mu\nu}{\cal O}^{\mu\nu,\alpha\beta}h_{\alpha\beta}$~\cite{Accioly:2017rft}.
Inverting ${\cal O}$, one obtains the propagator for the fourth-derivative gravity
\begin{equation} \label{propto10}
{\cal D}^{\rm 4th}_{\mu\nu,\alpha\beta}(k)=\Big[\frac{1}{k^2}-\frac{1}{k^2-m^2_2}\Big]P^{(2)}-\frac{1}{2}\Big[\frac{1}{k^2}-\frac{1}{k^2-m^2_0}\Big]P^{(0-s)}+(\cdots),
\end{equation}
where $P^{(2)}$ and $P^{(0-s)}$ represent the Barnes-Rivers operators,
\begin{eqnarray} \label{propto11}
P^{(2)}_{\mu\nu,\alpha\beta}&=&\frac{1}{2}(\theta_{\mu\alpha}\theta_{\nu\beta}+\theta_{\mu\beta}\theta_{\nu\alpha})-\frac{1}{3}\theta_{\mu\nu}\theta_{\alpha\beta},\\
P^{(0-s)}_{\mu\nu,\alpha\beta}&=&\frac{1}{3}\theta_{\mu\nu}\theta_{\alpha\beta},~~\theta_{\mu\nu}=\eta_{\mu\nu}-\frac{k_\mu k_\nu}{k^2},\label{propto12}
\end{eqnarray}
while $(\cdots)$ denotes the set of terms that are irrelevant to the spectrum of the theory.
Here  the spin-2 and spin-0 mass squared are given by, respectively,
\begin{equation}\label{mass1}
  m^2_2=-\frac{4}{\beta \kappa^2},~~~m^2_0=\frac{2}{\kappa^2(3\alpha+\beta)}.
\end{equation}
We require $\beta<0$ and $3\alpha+\beta>0$ for having non-tachyonic masses. The propagator (\ref{propto10}) carries  8 DOF: massless tensor (2 DOF) $+$ massive tensor (5 DOF) $+$ massless scalar (1 DOF).
We would like  to note  that (\ref{propto10}) without $(\cdots)$ represents  a gauge-invariant part of the propagator.
At this stage, it is worth noting that for large momentum, we have $\theta_{\mu\nu}\sim 1$ which implies that the  power counting argument is valid here because the massive tensor propagator takes the form of $\frac{P^{(2)}(\theta)}{k^2-m^2_2}\sim \frac{1}{k^2}$ . However, we have $\frac{P^{(2)}(\tilde{\theta})}{k^2-m^2}\sim \frac{k^2}{m^2}$ for the FP massive gravity with $\tilde{\theta}_{\mu\nu}=\eta_{\mu\nu}-k_\mu k_\nu/m^2$~\cite{Hinterbichler:2011tt}  which implies that the power counting argument does  not work here and thus, one cannot deduce the renormalizability of the FP massive gravity~\cite{Weinberg:1959nj}.

The spatial part
of the gauge-invariant propagator (\ref{propto10})  takes the form
\begin{eqnarray}
{\cal P}^{\rm 4th}_{\mu\nu,\alpha\beta}(\mathbf{k})=&-&\frac{1}{\mathbf{k}^2}\Big[\frac{1}{2}(\eta_{\mu\kappa}\eta_{\nu\lambda}+\eta_{\mu\lambda}\eta_{\nu\kappa})-\frac{1}{2}\eta_{\mu\nu}\eta_{\kappa\lambda}\Big]\nonumber \\
&+& \frac{1}{\mathbf{k}^2+m^2_2}\Big[\frac{1}{2}(\eta_{\mu\kappa}\eta_{\nu\lambda}+\eta_{\mu\lambda}\eta_{\nu\kappa})-\frac{1}{3}\eta_{\mu\nu}\eta_{\kappa\lambda}\Big] \label{propa1} \\
&-& \frac{1}{\mathbf{k}^2+m^2_0}\frac{ \eta_{\mu\nu}\eta_{\kappa\lambda}}{6} \nonumber,
\end{eqnarray}
where the second coefficient  $1/2(=1/3+1/6)$ in the first line differs  from  1/3 in the second line.
We note the relation between the Newtonian potential sourced by a static  mass $M$ and propagator
\begin{equation} \label{pp-rel}
V(r)=\frac{\kappa_4M}{(2\pi)^3}\int d^3\mathbf{k} e^{i\mathbf{ k}\cdot\mathbf{ r}} {\cal P}_{00,00}(\mathbf{k}).
\end{equation}
Fourier-transforming
\begin{equation}\label{propa1z}
  {\cal P}^{\rm 4th}_{00,00}(\mathbf{k})=\frac{1}{2}\Big[-\frac{1}{\mathbf{k}^2}+\frac{4}{3}\frac{1}{\mathbf{k}^2+m^2_2}-\frac{1}{3}\frac{1}{\mathbf{k}^2+m^2_0}\Big]
\end{equation}
 leads to the Newtonian potential   as
\begin{equation} \label{pot1}
V^{\rm 4th}(r)= \frac{GM}{r}\Big[-1+\frac{4}{3}e^{-m_2 r}-\frac{1}{3}e^{-m_0 r}\Big],
\end{equation}
which  was already deduced by the Stelle's work~\cite{Stelle:1976gc}.

Here we point out that in the limit of $r\to 0$, a massive ghost tensor contributes  $4/3(=1+1/3)$ to the Newtonian potential
 and  a massive healthy scalar contributes $- 1/3$ to the potential.  The singularity cancellation occurs in the fourth-derivative gravity.
  This model shows that the theory without any kind of nonlocality could be free from the Newtonian singularity.

   We need to explain the singularity cancellation by introducing a different mechanism
   instead of taking $r=0$ limit.
  First of all,  we wish to explain the appearance of  ``4/3" explicitly.
   In the zero mass  limit of $m_2\to 0(e^{-m_2r}\approx 1$), one has  not 1 but 4/3 zeroth order term which indicates that the vDVZ discontinuity occurs  in the
  linearized fourth-derivative gravity. We could not distinguish the zero mass limit from the $r\to 0$ limit, because two cases provide the same zeroth order term of $e^{-m_2r}\approx 1$ in the Newtonian potential.
  However, one has to focus on the zero mass limit to introduce  the other  mechanism of the vDVZ discontinuity.
    The vDVZ discontinuity  dictates   that  the gravity theory obtained from massive one with 5 DOF by taking zero mass limit is not equivalent to the gravity theory obtained in the zero mass case.
    Especially, the former has 3 DOF, while the latter has 2 DOF which describes a massless tensor.
    A physical explanation of this phenomenon is that a massive tensor with mass $m$ carries 5 polarizations, while a massless tensor  carries only two~\cite{Porrati:2002cp}.  In the zero mass limit of $m\to0$, a massive tensor  decomposes into massless fields
of spin-2, 1 and 0.  The spin-0 field couples to the
trace of the stress-energy tensor. Therefore,  in the zero mass limit,  one does not recover the Einstein  gravity but rather a scalar-tensor theory.

  The discontinuity can be easily found by noting the difference in coefficients between the massless tensor propagator  (the first line) and massive ghost tensor one (the second line) in Eq. (\ref{propa1}).
 In the former we have 1/2, whereas we have 1/3 in the latter.  Thus, the zero mass limit of the massive propagator does not coincide with the massless propagator.

It is worth noting that   surviving DOF  in the zero mass limit is  3 DOF (one DOF is represented by an extra ghost scalar and the other 2 DOF are given by a  ghost tensor).
This massive ghost scalar with 1/3 could be   identified with the Stueckelberg scalar~\cite{Hinterbichler:2011tt} and it cancels against a massive healthy scalar with $-1/3$. On the other hand,  a ghost tensor with 1 cancels out a healthy tensor (Newton  term with $-1$).
The  vDVZ discontinuity explains why ${\cal O}(1)/r$ disappeared well.  Hence, we may  avoid  the singularity at the origin.

Consequently, the cancellation of singularity occurs because there is contribution ($-1,4/3,-1/3$) from 6 DOF [massless tensor with $2^+$, massless limit of massive ghost  tensor with $3^-(=2^-+1^-)$, and
massive scalar with $1^+$, where the superscripts $+(-)$ represent healthy (ghost) DOF].  Considering 8 DOF of the  theory initially, it is clear that the vDVZ discontinuity occurs in the zero mass  limit of a massive ghost tensor.

\section{Full sixth-derivative theory of gravity}
An action for full  sixth-derivative gravity takes the form
 \begin{equation} \label{action2}
S_{\rm 6th}=\int d^4x \sqrt{-g}\frac{1}{\kappa^2}\Big[2 R+\frac{\alpha_0}{2}R^2+\frac{\beta_0}{2}R_{\mu\nu}^2+\frac{\alpha_1}{2}R\square R+\frac{\beta_1}{2}R_{\mu\nu}\square R^{\mu\nu}\Big].
 \end{equation}
 It was proven that the action (\ref{action2}) becomes  super-renormalizable because the superficial divergence $\delta[=4+E-\sum_{n=3}^{\infty}(n-2)V_n]$ decreases as the number of vertices $V_n$ increases~\cite{Asorey:1996hz,Accioly:2017rft}. An  important point to be reminded is that the same order of the last two six-derivative terms in (\ref{action2}) is a key to guarantee the renormalizability.
 The propagator is found to be
 \begin{eqnarray}
  {\cal D}^{\rm 6th}(k)&=&\Big[\frac{1}{k^2}+\frac{1}{m^2_{2_+}-m^2_{2_-}}\Big(\frac{m^2_{2_-}}{k^2-m^2_{2_+}}-\frac{m^2_{2_+}}{k^2-m^2_{2_-}}\Big)\Big]P^{(2)} \nonumber  \\
 &-&\frac{1}{2}\Big[\frac{1}{k^2}+\frac{1}{m^2_{0_+}-m^2_{0_-}}\Big(\frac{m^2_{0_-}}{k^2-m^2_{0_+}}-\frac{m^2_{0_+}}{k^2-m^2_{0_-}}\Big)\Big]P^{(0-s)} \label{propa30z}\\
 &+&(\cdots)\nonumber,
\end{eqnarray}
where mass squared $m^2_{2\pm}$ and  $m^2_{0\pm}$ are defined by
\begin{eqnarray}
m^2_{2\pm}&=&\frac{\beta_0}{2\beta_1}\Big(1\pm \sqrt{1+\frac{16\beta_1}{\beta_0^2}}\Big), \nonumber \\
m^2_{0\pm}&=&\frac{3\alpha_0+\beta_0}{2(3\alpha_1+\beta_1)}\Big(1\pm \sqrt{1-\frac{8(3\alpha_1+\beta_1)}{(3\alpha_0+\beta_0)^2}}\Big).\nonumber
\end{eqnarray}
Here one requires $\beta_0<0$ and $\beta_1<1$ to have non-tachyonic masses. In this case, $ {\cal D}^{\rm 6th}(k)$ describes propagations of 14 (=2+5+5+1+1) DOF.
Its spatial part of the gauge-invariant propagator is given by
\begin{eqnarray}
  {\cal P}^{\rm 6th}_{00,00}(\mathbf{k})=\frac{1}{2}\Big[&-&\frac{1}{\mathbf{k}^2}+\frac{4}{3}\frac{1}{m^2_{2_+}-m^2_{2_-}}\Big(\frac{m^2_{2_+}}{\mathbf{k}^2+m^2_{2_-}}-\frac{m^2_{2_-}}{\mathbf{k}^2+m^2_{2_+}}\Big)\nonumber\\
 &-&\frac{1}{3}\frac{1}{m^2_{0_+}-m^2_{0_-}}\Big(\frac{m^2_{0_+}}{\mathbf{k}^2+m^2_{0_-}}-\frac{m^2_{0_-}}{\mathbf{k}^2+m^2_{0_+}}\Big)\Big]. \label{propa3z}
\end{eqnarray}
The particle content of the model is made up of three healthy particles (massless tensor, massive tensor with mass $m_{2+}$, massive scalar with $m_{0-}$)
and two ghosts (massive tensor with $m_{2-}$ and massive scalar with $m_{0+}$).

In this case, the Newtonian potential generated by static mass $M$ is  derived to be
\begin{eqnarray}
V^{\rm 6th}(r)= \frac{GM}{r}\Big[&-&1+\frac{4}{3}\frac{m^2_{2_+}e^{-m_{2_-} r}-m^2_{2_-}e^{-m_{2_+} r}}{m^2_{2_+}-m^2_{2_-}} \nonumber \\
&-&\frac{1}{3}\frac{m^2_{0_+}e^{-m_{0_-} r}-m^2_{0_-}e^{-m_{0_+} r}}{m^2_{0_+}-m^2_{0_-}}\Big]. \label{pot2}
\end{eqnarray}
The massive particle content  is made by taking the small mass limit ($e^{-m_i r} \approx 1$) of  massive ghost and healthy tensors  with 6 DOF
\begin{equation}
 \frac{4}{3}\Big\{ \frac{m^2_{2_+}}{m^2_{2_+}-m^2_{2_-}},-\frac{m^2_{2_-}}{m^2_{2_+}-m^2_{2_-}}\Big\} \Rightarrow \frac{4}{3}=1+\frac{1}{3},
\end{equation}
which provides 4/3 in the zeroth  order amplitude. Here we call the zero mass limit as the small mass limit because the masses are not zero but they  are so small that one can take  $e^{-m_i r} \approx 1$.
Also,  the  massive healthy and ghost  scalars  with 2 DOF  provide
\begin{equation}
-\frac{1}{3}\Big\{\frac{m^2_{0_+}}{m^2_{0_+}-m^2_{0_-}},-\frac{m^2_{0_-}}{m^2_{0_+}-m^2_{0_-}}\Big\}  \Rightarrow -\frac{1}{3},
\end{equation}
which provides $-1/3$ in the zeroth order amplitude.
Considering 14 DOF of the theory, we have 10 $(=2^++3^-+3^++1^++1^-)$ DOF in the Newtonian potential.
This shows clearly that  the vDVZ discontinuity occurs in the small mass limit of massive ghost and healthy tensors.
We note that the singularity cancellation in $V_{\rm 6th}(r)$  occurs either in the small mass limit or the $r\to0$ limit
which gives the same zeroth order  approximation of $e^{-m^2_ir}\approx1$ for $i=2_{\pm},0_{\pm}$.

This model  gives an affirmative answer to   the conjecture that the cancellation mechanism of the singularity is
the matching of the ghost and healthy modes between the spin-2 and spin-0 massive sectors: $(3^-,3^+)$ and $(1^+,1^-)$.

\section{Half sixth-derivative theory of gravity}
It was proposed that  if a higher-derivative gravity is renormalizable, it implies  necessarily  a finite Newtonian potential at the origin,
but the reverse of this statement is not true. A relevant action is given by the half six-derivative gravity as~\cite{Giacchini:2016xns,Accioly:2017rft,Quandt:1990gc}
\begin{equation} \label{action3}
S_{\rm h6th}=\int d^4x \sqrt{-g}\Big[\frac{2}{\kappa^2} R+a_0R^2+a_1 R\square R+b_0R_{\mu\nu}^2\Big].
 \end{equation}
 This action is not renormalizable because different derivative orders loose renormalizability~\cite{Asorey:1996hz}. A matching factor  of sixth-derivative term $R_{\mu\nu}\square R^{\mu\nu}$ was missed
 in the action (\ref{action3}).
 Its propagator takes the form
 \begin{eqnarray}
  {\cal D}^{\rm h6th}(k)&=&\Big[\frac{1}{k^2}+\frac{1}{k^2-\tilde{m}^2_{2}}\Big]P^{(2)} \nonumber  \\
 &-&\frac{1}{2}\Big[\frac{1}{k^2}+\frac{1}{\tilde{m}^2_{0_+}-\tilde{m}^2_{0_-}}\Big(\frac{\tilde{m}^2_{0_-}}{k^2-\tilde{m}^2_{0_+}}-\frac{\tilde{m}^2_{0_+}}{k^2-\tilde{m}^2_{0_-}}\Big)\Big]P^{(0-s)} \label{propa40z}\\
 &+&(\cdots), \nonumber
\end{eqnarray}
where mass squared $\tilde{m}^2_{2}$ and  $\tilde{m}^2_{0\pm}$ are defined by
\begin{eqnarray}
\tilde{m}^2_{2}&=&-\frac{4}{b_0 \kappa^2}, \nonumber \\
\tilde{m}^2_{0\pm}&=&\frac{3a_0+b_0\pm \sqrt{(3a_0+b_0)^2-24a_1/\kappa^2}}{6a_1}.\nonumber
\end{eqnarray}
The propagator (\ref{propa40z})  describes 9 $(=2+5+1+1)$ DOF of the theory.
In this case, the (00,00)-spatial part of the propagator is given by
\begin{eqnarray}
\label{propa3z}
  {\cal P}^{\rm h6th}_{00,00}(\mathbf{k})=\frac{1}{2}\Big[&-&\frac{1}{\mathbf{k}^2}+\frac{4}{3}\frac{1}{\mathbf{k}^2+\tilde{m}^2_2} \nonumber \\
 &-&\frac{1}{3}\frac{1}{\tilde{m}^2_{0_+}-\tilde{m}^2_{0_-}}\Big(\frac{\tilde{m}^2_{0_+}}{\mathbf{k}^2+\tilde{m}^2_{0_-}}-\frac{\tilde{m}^2_{0_-}}{\mathbf{k}^2+\tilde{m}^2_{0_+}}\Big)\Big]. \label{propa2z}
\end{eqnarray}
The potential is given by
\begin{equation} \label{pot3}
V^{\rm h6th}(r)= \frac{GM}{r}\Big[-1+\frac{4}{3}e^{-\tilde{m}_2 r}-\frac{1}{3}\frac{\tilde{m}^2_{0_+}e^{-\tilde{m}_{0_-} r}-\tilde{m}^2_{0_-}e^{-\tilde{m}_{0_+} r}}{\tilde{m}^2_{0_+}-\tilde{m}^2_{0_-}}\Big].
\end{equation}
The singularity is cancelled despite the fact that there is no massive healthy tensor to balance a massive ghost scalar.
It seems that this  case give a negative answer to the conjecture that the cancellation mechanism of the singularity requires  the matching of the ghost and healthy modes between the spin-2 and spin-0 massive sectors.
Even though the cancellation of singularity occurs in the $r\to 0$ limit of $e^{-\tilde{m}_{i} r}\approx1$ for $i=2,0_{\pm}$, this is not the case.
Considering 9 DOF of the theory, we have 7 $(=2^++3^-+1^++1^-)$ DOF  in the Newtonian potential.
Here  the vDVZ discontinuity occurs only  in the small mass limit of a massive ghost tensor, leaving a mismatch for a massive ghost scalar with $1^-$.
The number of massive excitations in each sector should  be the same. That is,  there   exists a massive healthy tensor  to each ghost mode in scalar sector and vice versa.
However, we have a particle content of ($3^-,\bullet$) and ($1^+,1^-$).
In this case, a massive healthy tensor with 5 DOF is necessary to make  a  renormalizable theory like the full sixth-derivative gravity (\ref{action2}) and to make a balance with a massive ghost scalar.
Then, $\bullet$ is given by $3^+$.

Similarly, we propose that  the other half six-derivative gravity
\begin{equation} \label{action4}
\tilde{S}_{\rm h6th}=\int d^4x \sqrt{-g}\Big[\frac{2}{\kappa^2} R+a_0R^2+b_0R_{\mu\nu}^2+b_1R_{\mu\nu}\square R^{\mu\nu}\Big].
 \end{equation}
is not renormalizable but it has a finite Newtonian potential at the origin. This model provides a particle content of ($3^-,3^+$) and ($1^+,\bullet$).
Here a massive ghost scalar ($\bullet$)  is needed to make a renormalizable theory like the full sixth-derivative gravity (\ref{action2}).

\section{Infinite-derivative gravity}
In this section, we wish to comment on the connection between  a non-singular Newtonian potential  and the  vDVZ discontinuity in infinite-derivative gravity (IDG)~\cite{Modesto:2011kw,Biswas:2011ar}.
It was shown that the infinite-derivative gravity has provided a finite Newtonian  at the origin~\cite{Biswas:2011ar,Frolov:2015usa,Edholm:2016hbt,Conroy:2017nkc}.

 A simplest model of  IDG is given by~\cite{Modesto:2014eta}
\begin{equation} \label{idg-1}
S_{\rm IDG}=-\frac{1}{2\kappa^2}\int d^4x\sqrt{-g} \Big[R+G_{\mu\nu}\frac{a(\square_\Lambda)-1}{\square} R^{\mu\nu}\Big].
\end{equation}
The propagator of this IDG takes the form
\begin{equation}
{\cal D}^{\rm IDG}(k)=\frac{1}{k^2 a(-k^2/\Lambda^2)}\Big[P^{(2)}-\frac{P^{(0-s)}}{2}\Big],
\end{equation}
which shows a ghost-free propagator of  a massless tensor. If one chooses $a(\square_\Lambda)=e^{-\square/\Lambda^2}$~\cite{Tseytlin:1995uq}, there is no room to introduce masses of massive tensors. Its potential is found to be
\begin{equation}
V^{\rm IDG}(r)=-\frac{GM}{r}{\rm Erf}\Big(\frac{\Lambda r}{2}\Big),
\end{equation}
which is a non-singular potential at the origin because the error function takes the form of ${\rm Erf}(x)\sim 2x/\sqrt{\pi}$ as $x\to 0$.
Here, the error function is defined through (\ref{pp-rel}) by
\begin{equation} \label{newtp-2}
  {\rm  Erf} \Big(\frac{\Lambda r}{2}\Big)= \frac{2}{\pi}\int^{\infty}_{0}d|{\bf k}|\frac{e^{-|{\bf k}|^2/\Lambda^2} \sin[|{\bf k}|r]}{|{\bf k}|}.
 \end{equation}
 It is important to note   that although one does not require the small mass limit, the singularity disappears due to the non-locality
and the effect depends on a  form of $a(\square)$.  Hence, it seems that  the singularity cancellation has nothing to do with  the vDVZ discontinuity.
 Interestingly, the super-renormalizable and nonlocal massive gravity has provided a massive propagator~\cite{Modesto:2013jea}
\begin{equation}
{\cal D}^{\rm SN}(k)=\frac{ e^{-H(k^2/M^2)}}{k^2-m^2}\Big[P^{(2)}-\frac{P^{(0-s)}}{2}+\xi\Big(P^{(1)}+\frac{\tilde{P}^{(0)}}{2}\Big)\Big],
\end{equation}
which is the same form as that of a massless tensor except an overall factor. If one takes the zero mass limit of $m\to 0$, the massive propagator reduces smoothly to the massless one which shows
that there is no  vDVZ discontinuity.

To study the connection between  a finite Newtonian potential  and the  vDVZ discontinuity,  we consider a polynomial action  of IDG defined by~\cite{Modesto:2014eta}
\begin{equation}
\tilde{S}^{\rm IDG}=\frac{1}{\kappa^2}\int d^4x \sqrt{-g}\Big[-2R+RF_1(\square)R+R_{\mu\nu}F_2(\square) R^{\mu\nu}\Big], \label{idg-2}
\end{equation}
where
\begin{eqnarray}
&& F_1(\square)=\alpha_0+\alpha_1\square+\cdots +\alpha_N\square^N,\\
&& F_2(\square)=\beta_0+\beta_1\square+\cdots +\beta_N\square^N.
\end{eqnarray}
The $N=0[N=1]$ model corresponds to the fourth-derivative gravity (\ref{action1}) [sixth-derivative gravity (\ref{action2})] with different coefficients.
It requires that the two polynomials of $F_1$ and $F_2$ be of the same order.
A  potential generated by a static mass $M$ can be expressed as
\begin{equation} \label{idgpot}
\tilde{V}^{\rm IDG}(r)=-\frac{2GM}{\pi r}\int_{0}^{\infty}\frac{dk}{k}\sin[kr]\Big[\frac{4}{3}\frac{1}{P_{2N+2}(k)}-\frac{1}{3}\frac{1}{Q_{2N+2}(k)}\Big],
\end{equation}
where $P_{2N+2}(k)$ and $Q_{2N+2}(k)$ are polynomials of spin-2 and spin-0 massive sectors  given by
\begin{eqnarray}
P_{2N+2}(k)=1&+&\frac{1}{2}\Big[\beta_0k^2-\beta_1k^4+\cdots+(-1)^N\beta_N k^{2N+2}\Big], \\
Q_{2N+2}(k)=1&-&(3\alpha_0+\beta_0)k^2+(3\alpha_1+\beta_1)k^4+\cdots \nonumber \\
&+&(-1)^N(3\alpha_N+\beta_N)k^{2N+2}.
\end{eqnarray}
Factorizing $P_{2N+2}$ and $Q_{2N+2}$, one introduces masses of spin-0 and spin-2 massive sectors to have all simple poles as
\begin{equation}
0<m^2_{(k)0}<m^2_{(k)1}<\cdots <m^2_{(k)N}~{\rm and}~m^2_{(k)i}\not=m^2_{(k)j},~i\not=j
\end{equation}
for $k=0,2$. After contour integration, one arrives at the potential
\begin{eqnarray}
\tilde{V}^{\rm IDG}(r)=-\frac{GM}{r}\Big[1&-&\frac{4}{3}\sum_{i=0}^{N}\prod_{j\not=i}\frac{m^2_{(2)j}}{m^2_{(2)j}-m^2_{(2)i}}e^{-m_{(2)i}r} \nonumber \\
&+&\frac{1}{3}\sum_{i=0}^{N}\prod_{j\not=i}\frac{m^2_{(0)j}}{m^2_{(0)j}-m^2_{(0)i}}e^{-m_{(0)i}r}\Big] \label{p-idg}.
\end{eqnarray}
In the small mass limit of $e^{-m_{(2)i}r}\approx1$ and $e^{-m_{(0)i}r}\approx1$, one finds $[\cdots]$ in (\ref{p-idg})  as
\begin{equation}
-\frac{4}{3}\sum_{i=0}^{N}\prod_{j\not=i}\frac{m^2_{(2)j}}{m^2_{(2)j}-m^2_{(2)i}}+\frac{1}{3}\sum_{i=0}^{N}\prod_{j\not=i}\frac{m^2_{(0)j}}{m^2_{(0)j}-m^2_{(0)i}}.
\end{equation}
Considering the relation which is valid for any set of numbers $a_j$
\begin{equation}
\sum_{i=0}^{N}\prod_{j\not=i}\frac{a_j}{a_j-a_i}=1,
\end{equation}
we find that the sum of  zeroth order terms is zero
\begin{equation}
-1+\frac{4}{3}-\frac{1}{3}=0. \label{lcoef}
\end{equation}
Considering the propagator in the integrand of (\ref{idgpot}), the  total DOF is $2+12N(=2+10N+2N)$. However, we have $2+8N(=2^++(3N)^- +(3N)^++N^+ +N^-$) DOF in the Newtonian potential (\ref{p-idg}).
This shows clearly that the vDVZ discontinuity occurs in the small mass limit of spin-2 massive sector.
Here we decompose $(3N)^-$ into $(N+2N)^-$ in the spin-2 massive sector where $N^-$ is represented by $N$ ghost Stueckelberg scalars, while $(2N)^-$ by $N$ massless ghost tensors. This happens to  $(3N)^+$ for massive healthy tensors, similarly.
They are  equivalent to write $4/3$ in (\ref{lcoef})  to be $1/3+1$.
The last two $N^+(N^-)$ can be represented by $N$ healthy (ghost) massive scalars, providing $-1/3$ in (\ref{lcoef}).
We have a particle content of $[(3N)^-,(3N)^+]$ and $[(N)^-,(N)^+]$.
It explains why the matching of the number of  healthy and ghost  modes between the spin-2 and spin-0 massive sectors is essential to make the singularity cancellation in the Newtonian potential.

Finally, this might correspond to the condition of super-renormalizability~\cite{Modesto:2014eta,Accioly:2017xmm}.
At this point, it would be better to distinguish three types of renormalizable theory.
(i) Finite: no counterterms need at all. (ii) Super-renormalizable: only a finite number of graphs need overall counterterms.
(iii) Renormalizable: infinitely many number of graphs need overall counterterms.
 (But note that they only normalize a finite set of terms  in the basic Lagrangian since we assumed renormalizability of the theory.)

\section{Discussions}

First of all, we have shown that  the vDVZ discontinuity is  related closely  to the  the singularity cancellation of the Newtonian potential.
For this purpose, we have chosen the zero (small) mass limit of $m_i\to 0$ instead of the $r\to 0$ limit.

In this work, we have  explored why the matching of the number of ghost and  healthy  modes between the spin-2 and spin-0 massive sectors is necessary to make the singularity cancellation.
This was  explained by introducing the vDVZ discontinuity appeared  in the zero mass limit  of higher-derivative gravity.
Therefore,  if a higher-derivative gravity is renormalizable, it implies  necessarily  a finite Newtonian potential at the origin.
Furthermore, the reverse of this statement seems to be  true. Although  a counter example of (\ref{action3}) which is not renormalizable  provides a finite potential at the origin, the vDVZ discontinuity occurs only in the small mass limit of a massive ghost tensor. In this case, a massive healthy tensor is needed to make a renormalizable theory which amounts to happening that  the vDVZ discontinuity occurs in the small mass limit of both  massive ghost and healthy tensors. This leads  to a balance between the attractive forces and repulsive forces in each sector as well as a specific matching of the number of tensor and scalar modes.

We note that the effect of singularity cancellation and the vDVZ discontinuity are  linear effects involving the independent contribution of scalars and tensors.
Hence, it might not be clear that the cancellation may hold in these theories at the nonlinear level.
However, it seems that  the UV divergences of quantum theory are related to the Newtonian singularity.
This  means  that the Newtonian singularity is indeed the simplest UV divergence due to the interaction.
Also, we mention that two nonlinear issues of Vainshtein radius~\cite{Vainshtein:1972sx} and Boulware-Deser ghost~\cite{Boulware:1973my} concerning  the  vDVZ discontinuity are not directly related to the renormalizability.

On the other hand, the other physical observable of light deflection (bending angle) does not depend on the massive spin-0 sector of $R^2$ and $R\square R$~\cite{Accioly:2001cc,Accioly:2016lzp}.
Thus, it suggests  that  the UV divergence of quantum theory is  not closely  related to the light bending angle.

Finally, the two tracks were found  to arrive at a finite Newtonian at the origin. One is to use the IDG action (\ref{idg-1}) without ghost fields. In this case, the singularity disappeared due to the nonlocality.
The other is to consider the IDG action (\ref{idg-2}) with ghost fields. The ghost scalar and tensors are needed  to have a finite potential at the origin.

\section*{Acknowledgement}
This work was supported by the National Research Foundation of Korea (NRF) grant funded by the Korea government (MOE) (No. NRF-2017R1A2B4002057).


\newpage
\begin{thebibliography}{99}
\bibitem{Accioly:2013hwa}
  A.~Accioly, J.~Helayel-Neto, E.~Scatena and R.~Turcati,
  Int.\ J.\ Mod.\ Phys.\ D {\bf 22}, 1342015 (2013).
  doi:10.1142/S0218271813420157


\bibitem{Modesto:2014eta}
  L.~Modesto, T.~de Paula Netto and I.~L.~Shapiro,
  JHEP {\bf 1504}, 098 (2015)
  doi:10.1007/JHEP04(2015)098
  [arXiv:1412.0740 [hep-th]].

\bibitem{Stelle:1976gc}
  K.~S.~Stelle,
  Phys.\ Rev.\ D {\bf 16}, 953 (1977).
  doi:10.1103/PhysRevD.16.953

\bibitem{Asorey:1996hz}
  M.~Asorey, J.~L.~Lopez and I.~L.~Shapiro,
  Int.\ J.\ Mod.\ Phys.\ A {\bf 12}, 5711 (1997)
  doi:10.1142/S0217751X97002991
  [hep-th/9610006].


\bibitem{Giacchini:2016xns}
  B.~L.~Giacchini,
  Phys.\ Lett.\ B {\bf 766}, 306 (2017)
  doi:10.1016/j.physletb.2017.01.019
  [arXiv:1609.05432 [hep-th]].

\bibitem{Accioly:2017rft}
  A.~Accioly, J.~de Almeida, G.~P.~Brito and G.~Correia,
  Phys.\ Rev.\ D {\bf 95}, no. 8, 084007 (2017)
  doi:10.1103/PhysRevD.95.084007
  [arXiv:1702.07404 [hep-th]].


\bibitem{Fierz:1939ix}
  M.~Fierz and W.~Pauli,
  Proc.\ Roy.\ Soc.\ Lond.\ A {\bf 173}, 211 (1939).
  doi:10.1098/rspa.1939.0140

\bibitem{vanDam:1970vg}
  H.~van Dam and M.~J.~G.~Veltman,
  Nucl.\ Phys.\ B {\bf 22}, 397 (1970).
  doi:10.1016/0550-3213(70)90416-5

\bibitem{Zakharov:1970cc}
  V.~I.~Zakharov,
  JETP Lett.\  {\bf 12}, 312 (1970)
  [Pisma Zh.\ Eksp.\ Teor.\ Fiz.\  {\bf 12}, 447 (1970)].

\bibitem{Stueckelberg:1957zz}
  E.~C.~G.~Stueckelberg,
  Helv.\ Phys.\ Acta {\bf 30}, 209 (1957).


\bibitem{Hinterbichler:2011tt}
  K.~Hinterbichler,
  Rev.\ Mod.\ Phys.\  {\bf 84}, 671 (2012)
  doi:10.1103/RevModPhys.84.671
  [arXiv:1105.3735 [hep-th]].

\bibitem{Weinberg:1959nj}
  S.~Weinberg,
  Phys.\ Rev.\  {\bf 118}, 838 (1960).
  doi:10.1103/PhysRev.118.838

\bibitem{Porrati:2002cp}
  M.~Porrati,
  Phys.\ Lett.\ B {\bf 534}, 209 (2002)
  doi:10.1016/S0370-2693(02)01656-8
  [hep-th/0203014].



\bibitem{Quandt:1990gc}
  I.~Quandt and H.~J.~Schmidt,
  Astron.\ Nachr.\  {\bf 312}, 97 (1991)
  doi:10.1002/asna.2113120205
  [gr-qc/0109005].


\bibitem{Modesto:2011kw}
  L.~Modesto,
  Phys.\ Rev.\ D {\bf 86}, 044005 (2012)
  doi:10.1103/PhysRevD.86.044005
  [arXiv:1107.2403 [hep-th]].


\bibitem{Biswas:2011ar}
  T.~Biswas, E.~Gerwick, T.~Koivisto and A.~Mazumdar,
  Phys.\ Rev.\ Lett.\  {\bf 108}, 031101 (2012)
  doi:10.1103/PhysRevLett.108.031101
  [arXiv:1110.5249 [gr-qc]].

\bibitem{Frolov:2015usa}
  V.~P.~Frolov and A.~Zelnikov,
  Phys.\ Rev.\ D {\bf 93}, no. 6, 064048 (2016)
  doi:10.1103/PhysRevD.93.064048
  [arXiv:1509.03336 [hep-th]].

\bibitem{Edholm:2016hbt}
  J.~Edholm, A.~S.~Koshelev and A.~Mazumdar,
  Phys.\ Rev.\ D {\bf 94}, no. 10, 104033 (2016)
  doi:10.1103/PhysRevD.94.104033
  [arXiv:1604.01989 [gr-qc]].

\bibitem{Conroy:2017nkc}
  J.~Edholm and A.~Conroy,
  Phys.\ Rev.\ D {\bf 96}, no. 4, 044012 (2017)
  doi:10.1103/PhysRevD.96.044012
  [arXiv:1705.02382 [gr-qc]].

\bibitem{Tseytlin:1995uq}
  A.~A.~Tseytlin,
  Phys.\ Lett.\ B {\bf 363}, 223 (1995)
  doi:10.1016/0370-2693(95)01228-7
  [hep-th/9509050].

\bibitem{Modesto:2013jea}
  L.~Modesto and S.~Tsujikawa,
  Phys.\ Lett.\ B {\bf 727}, 48 (2013)
  doi:10.1016/j.physletb.2013.10.037
  [arXiv:1307.6968 [hep-th]].



\bibitem{Accioly:2017xmm}
  A.~Accioly, J.~de Almeida, G.~P.~de Brito and W.~Herdy,
  arXiv:1707.02083 [hep-th].

\bibitem{Vainshtein:1972sx}
  A.~I.~Vainshtein,
  Phys.\ Lett.\  {\bf 39B}, 393 (1972).
  doi:10.1016/0370-2693(72)90147-5


\bibitem{Boulware:1973my}
  D.~G.~Boulware and S.~Deser,
  Phys.\ Rev.\ D {\bf 6}, 3368 (1972).
  doi:10.1103/PhysRevD.6.3368

\bibitem{Accioly:2001cc}
  A.~Accioly and H.~Blas,
  Phys.\ Rev.\ D {\bf 64}, 067701 (2001)
  doi:10.1103/PhysRevD.64.067701
  [gr-qc/0107003].

\bibitem{Accioly:2016lzp}
  A.~Accioly, B.~L.~Giacchini and I.~L.~Shapiro,
  arXiv:1610.05856 [gr-qc].






 \end{thebibliography}
\end{document}